\documentclass[conference]{IEEEtran}
\IEEEoverridecommandlockouts
\usepackage{cite}
\usepackage{amsmath,amssymb,amsfonts}
\usepackage{algorithmic}
\usepackage{graphicx}
\usepackage{textcomp}
\usepackage{xcolor}
\usepackage{tikz}
\usepackage{balance}

\usepackage{pgfplots}
\usepackage{pgfplotstable}
\pgfplotsset{compat=1.7}
\usepgfplotslibrary{groupplots}
\usepackage{soul} 
\sethlcolor{orange}
\usepackage[normalem]{ulem} 
\usepackage[scientific-notation=true]{siunitx}

\usetikzlibrary{external}

\usepackage{amsmath}

\def\BibTeX{{\rm B\kern-.05em{\sc i\kern-.025em b}\kern-.08em
    T\kern-.1667em\lower.7ex\hbox{E}\kern-.125emX}}

\begin{document}

\title{Rateless Autoencoder Codes:\\ Trading off Decoding Delay and Reliability\\

\thanks{This paper has received funding from the European Union’s Horizon 2020 research and innovation programme under Grant Agreement number 856967.}
}

\author{\IEEEauthorblockN{Vukan Ninkovic,\IEEEauthorrefmark{1} Dejan Vukobratovic,\IEEEauthorrefmark{1} {Christian H\"{a}ger,\IEEEauthorrefmark{2} Henk Wymeersch,\IEEEauthorrefmark{2}   and Alexandre Graell i Amat\IEEEauthorrefmark{2}}
\IEEEauthorblockA{\IEEEauthorrefmark{1}University of Novi Sad, 
Novi Sad, Serbia}
\IEEEauthorblockA{\IEEEauthorrefmark{2}Department of Electrical Engineering, Chalmers University of Technology, Gothenburg, Sweden
}}}

\maketitle

\begin{abstract}
Most of today's communication systems are designed to target reliable message recovery after receiving the entire encoded message (codeword). However, in many practical scenarios, the transmission process may be interrupted before receiving the complete codeword. This paper proposes a novel rateless autoencoder (AE)-based code design suitable for decoding the transmitted message before the noisy codeword is fully received. 
Using particular dropout strategies applied during the training process, rateless AE codes allow to trade off   between decoding delay and reliability, providing a graceful improvement of the latter with each additionally received codeword symbol. The proposed rateless AEs  significantly outperform the conventional AE designs for scenarios where it is desirable to trade off reliability for lower decoding delay.

\end{abstract}


\IEEEpeerreviewmaketitle

\section{Introduction}
\label{intro}


The design of short block-length error-correcting codes for unpredictable and time-varying wireless channels still remains a challenge \cite{Shirvanimoghaddam_2018}. 
Particularly challenging is the design of codes for channels experiencing prolonged deep fades or even  complete channel failures that prevent the receiver to receive the complete (noisy) codeword. Such channels, referred to as \emph{dying channels} in \cite{Zeng_2008} and \cite{Varshney_2012}, arise in various communication systems, e.g., due to loss of synchronization, lack of memory, depletion of harvested energy in wireless sensors, interruption of a secondary user by a primary user in cognitive radio, loss of line-of-sight channel in optical wireless communications, limited channel duration of low-earth-orbit satellite communications, or physical defects in magnetic recording memories \cite{Varshney_2012,Gu_2017, Zeng_2012}. 
Despite their apparent applicability, few works have considered the design of coding schemes for dying channels \cite{Gu_2017, Zeng_2012, Zeng_2013}.


In the above--mentioned scenarios, a conventional fixed-rate code design  may suffer from high inefficiency in the short block-length regime. 
Rateless codes provide a possibility for the receiver to trade off decoding delay with  increased reliability by adaptively receiving additional codeword symbols (thus decreasing the code rate) until the desired reliability is attained \cite{Luby_2002}. Rateless codes such as rateless spinal codes \cite{Perry_2011} and analog fountain codes (AFC) \cite{Lim_2021} are recent classes of codes that represent a flexible solution for rate adaptation to unpredictable wireless channels. Both spinal and AFC codes map the set of input messages directly into codewords comprising sequences of real (or complex) symbols, thus effectively performing adaptive coding and modulation. 

Deep autoencoders (AEs) provide a new framework  to design codes for challenging\textemdash or unknown\textemdash channels  \cite{OShea_2017}. 
AE-based codes  have been studied in several settings, such as one-bit quantization channels \cite {Balevi_2020}, optical communications \cite{Li_2018}, and OFDM \cite{Felix_2018}. 
The resulting AE-based codes are shown to perform close to optimal for several baseline scenarios \cite{OShea_2017, Balevi_2020, Li_2018, Felix_2018, Dorner_2018}. Similar to spinal and AFC codes, AE-based codes map input messages directly into real (or complex) codeword sequences. However, AE-based codes are trained\textemdash and thus optimized\textemdash  for a given code rate, meaning that they perform poorly if the decoding is attempted before the complete codeword is received. 

Designing flexible and efficient short block-length codes for dying channels \cite{Varshney_2012} is the focus of this paper. In particular, we extend the conventional AE-based code design to a class of AE-based codes that shares  with rateless codes that receiving additional codeword symbols progressively improves the successful message decoding probability. We call the proposed AE-based codes \emph{rateless AE codes}. Such progressive improvement of the reliability  allows the receiver to select when, i.e., after how many received symbols, to attempt decoding, thus trading off  error probability with decoding delay. Inspired by a recent work on rateless AEs for flexible reduced-dimensionality signal representation \cite{Koike-Akino_2020}, we integrate suitably designed dropout strategies into the AE-based code design  to induce the desired performance behavior. More precisely, by controlling the dropout parameters in rateless AE code design, we shape a desired decoding delay vs reliability behavior of the resulting codes: a property that is not easy to enforce using classical coding approaches. Numerical results demonstrate that the resulting rateless AE codes significantly outperform conventional AE-based code designs for scenarios where it is desirable to trade off reliability for lower decoding delay such as the case of dying channels.



\section{System Model and Autoencoder-Based Codes}
\label{system_model}

\subsection{System Model}
\label{model}

We consider a transmitter that sends a message $m$ from a message set $\mathcal{M}=\{1,2,\ldots,M\}$ over a noisy channel. Each message is represented as a sequence of $k=\log_2(M)$ bits $\boldsymbol{s}=(s_1,s_2,\ldots,s_k)$. The encoder encodes a message $m$ into a transmitted codeword $\boldsymbol{x} = (x_1,x_2,\ldots,x_n)$ of length $n$ symbols. Formally, we define the encoder via the function $f:\mathcal{M} \rightarrow \mathbb{R}^n$. We consider two different power constraints for the codewords: i) a fixed power constraint, for which $\| \boldsymbol{x} \|_2^2=n$ holds for every $\boldsymbol{x}$, and ii) and average power constraint, for which $\frac{1}{M}\sum_{i=1}^{i=M} \|\boldsymbol{x}_i \|_2^2 = n$. 
The code rate is defined as $R=k/n$ [bits/channel use]. 

Let $\boldsymbol{y}=(y_1,y_2,\ldots,y_n) \in \mathbb{R}^n$ be the output of the channel (described in the next subsection). 
The decoder maps 
$\boldsymbol{y}$ into the estimated message $\hat{m}$ using the decoder transform $g: {\mathbb{R}}^n \rightarrow \mathcal{M}$.

\subsection{Channel Model}
\label{MC_SC}

We consider a channel model consisting of the cascade of an AWGN channel and an erasure channel. The motivation underpinning this model is to consider an AWGN channel affected by random interruptions or occasional \emph{deep fades}\textemdash modeled as erasures\textemdash whose locations are known to the receiver (see Section~\ref{Comments} below for more details). Let $E_b/N_0$ denote the energy per bit ($E_b$) to noise power spectral density ($N_0$) ratio.


We consider two variants of the proposed cascaded AWGN and erasure channel, as detailed below. In both models, the erasure channel is described by a set of $L$ channel states. We denote by $p_{\ell}$ the probability that the channel is in the $\ell$-th state. The erasure channel state distribution is defined as $\boldsymbol{p}=\{p_1,p_2,\ldots,p_L\}$, $\sum_{i=\ell}^L p_{\ell} = 1$. 
 
\textbf{Model 1: Channel Model with Tail Erasures.} In this model, the receiver receives the first $r_\ell$ symbols of $\boldsymbol{y}$ and the remaining $n-r_\ell$ symbols are erased. The $\ell$-th channel state is then defined by the pair $(p_{\ell},r_{\ell})$; if the erasure channel is at state $\ell$, the receiver receives $\boldsymbol{y}_{\ell}=\{y_1,y_2,\ldots,y_{r_{\ell}}\}$, while the remaining symbols are erased. Let $\boldsymbol{r}=\{r_1,r_2,\ldots,r_L\}$. The channel model with tail erasures is then defined by $\boldsymbol{p}$ and $\boldsymbol{r}$. 

\textbf{Model 2: Channel Model with Random Erasures.} In this model, the symbols of $\boldsymbol{y}$ are randomly erased. The $\ell$-th state of the erasure channel is defined by a pair $(p_{\ell},\epsilon_{\ell})$, where $\epsilon_{\ell} \in [0,1]$ is the symbol erasure probability. The  $L$  channel states are defined by $\boldsymbol{p}$ and the set of corresponding erasure probabilities $\boldsymbol{\epsilon}=\{\epsilon_1,\epsilon_2,\ldots,\epsilon_L\}$. 



\subsection{Comment on the  Channel Models}
\label{Comments}
 
Model 1 is known as a \emph{channel that dies}, and its information-theoretic properties are investigated in \cite{Varshney_2012}. The model is suitable for situations where the receiver, after receiving a certain (varying) number of symbols, can no longer receive additional symbols. This may be the case with low-cost devices due to, e.g., loss of synchronization, lack of memory, or depletion of harvested energy, in satellite communications, molecular communications, and in certain magnetic recording cases. The model is also motivated by a scenario where a receiver is able to trade off  decoding delay against error probability by performing decoding based only on the first received symbols: The error probability decreases with each received symbol at the expense of an increased delay. Depending on the error rate and delay requirements, the receiver can decide when to start the decoding process.  

Model 2 is a multi-state extension of a channel model, dubbed AWGN+erasure channel, considered in \cite{Sahai_2004}. This model, as well as the model with tail erasures, is also  suitable to model multicast to heterogeneous receivers. In this case, the erasure channel state probabilities model the fraction of receivers experiencing different erasure channel states. In our context, the model with random erasures is used as a reference for comparison with the tail erasure channel model.

\subsection{Problem Formulation}

Under the setup in Section~\ref{MC_SC}, the goal is to design a pair $(f,g)$ that, for a given number of received symbols,  minimizes the average message error probability
\begin{align}
P_{\textrm{e}} = \frac{1}{M} \sum_{m \in \mathcal{M}} \mathbb{P}\{\hat{m} \neq m|m\}\,.   
\end{align}
Before presenting the proposed rateless AE code design, we review the basics of the conventional AE-based codes.

\section{Preliminaries: Conventional Autoencoder-based Code Design}
\label{Conv_AE}

\begin{figure}[t]
	\centering
	\includegraphics[width=1\linewidth]{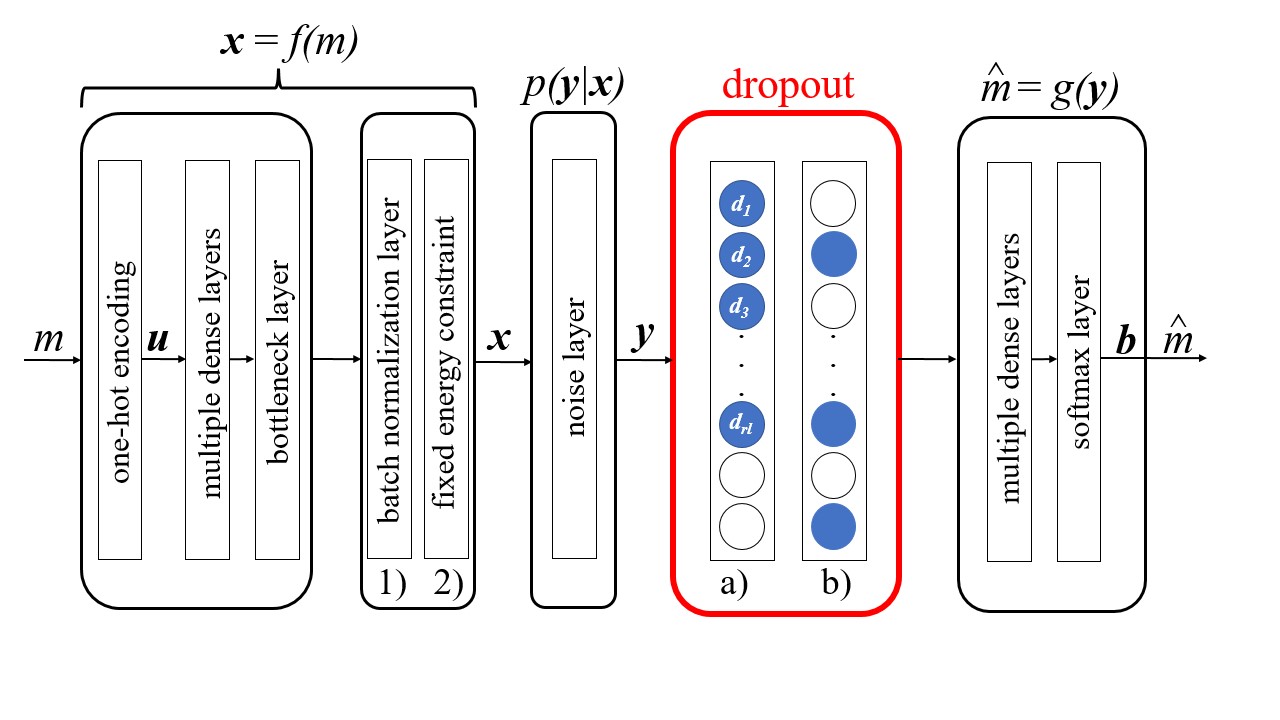}
	\vspace*{-7mm}
	\caption{Communication system represented as a deep autoencoder \cite{OShea_2017} with two types of normalization: 1) Average power constraint and 2) Fixed power constraint; and dropout block for: a) Model 1 and b) Model 2.}
	\label{Fig_AE}
\end{figure}

From the deep learning perspective,  the communication system described in Section~\ref{model} can be implemented as an autoencoder (AE) \cite{OShea_2017}, as illustrated in Fig.~\ref{Fig_AE} (disregard for the moment the dropout block). The input (encoder) and output (decoder) layers and the bottleneck layer constitute the main AE blocks. The AE-based approach introduces a new design paradigm in communication systems in which the transmitter and receiver components are jointly optimized using machine learning-based end-to-end learning methods.

 At the transmitter's input layer, the message $m$ is encoded as a one-hot vector $\boldsymbol{u} = (u_1,u_2,\ldots,u_M) \in \{0,1\}^M$, i.e., it is represented as an $M$-dimensional vector with the $m$-th element equal to one and all other elements equal to zero. The transmitter can be represented as a feedforward neural network with $H$ hidden layers, followed by a bottleneck layer of width $n$ (corresponding to the codeword length). At the output of the bottleneck layer, a normalization step  ensures that the power constraint on $\boldsymbol{x}$ is met.  The number of neurons in each hidden layer is given by the vector $\boldsymbol{h}=\{h_1,h_2,\ldots,h_H\}$, where $h_i$ represents the number of neurons of the $i$-th hidden layer.

 The AWGN channel is implemented by the noise layer.  The output of the noise layer can be represented as $\boldsymbol{y}=\boldsymbol{x}+\boldsymbol{z}$, where $\boldsymbol{z}$ contains $n$ independent and identically distributed  samples of a Gaussian random variable with zero mean and variance $\sigma^2$. It is important to notice that the random nature of the channel can be represented as a form of regularization, because the receiver never sees the same training example twice. As a consequence, it is almost impossible for the neural network to overfit \cite{Dorner_2018}. 

 The goal of the neural network is to find the most suitable representation of the information robust to the channel perturbations. The receiver is implemented in the same way as the transmitter (symmetric feedforward neural network), except that the last layer has a softmax activation function with output $\boldsymbol{b} = (b_1,b_2,\ldots,b_M) \in (0,1)^M$, $\Vert \boldsymbol{b}\Vert_1=1$. The index of the highest value element in $\boldsymbol{b}$ corresponds to the decoded message $\hat{m}$, i.e.,
 \begin{align*}
 \hat{m}=\arg\max_{i}\{b_i\}\,.  
\end{align*}
 
 Except for the last layer of the transmitter and receiver, which have linear and softmax activation functions respectively, all others  layers use  the rectified linear unit (ReLU) as the activation function. 
 
 Ideally, one would like to train the AE to minimize the error probability $P_{\textrm{e}}$. However, $P_{\textrm{e}}$ cannot be used directly as it is not differentiable. A common approach is to use the cross-entropy loss between $\boldsymbol{u}$ and $\boldsymbol{b}$,
  \begin{align}\label{eq2}
    \ell(\boldsymbol{u},\boldsymbol{b})=-\sum_{i=1}^{M} u_i \log{b_i}\,,    
    \end{align}
as a surrogate for the error probability. The AE is then trained so that  $\ell(\boldsymbol{u},\boldsymbol{b})$ is minimized.



\section{Rateless Autoencoder Codes}
\label{RA-design}

\subsection{Design of Rateless Autoencoder Codes}

We introduce a novel class of AE codes, referred to as  rateless AE codes, that allow to trade off decoding delay and reliability. Inspired by the rateless AEs in \cite{Koike-Akino_2020}, we  use a suitably defined \emph{randomized dropout} strategy  to match the AE-based code design to a given erasure channel model. To define a generic randomized dropout strategy covering various erasure channel models, we introduce a channel dropout vector $\boldsymbol{d}$ associated to the channel noise layers, as shown in Fig.\ref{Fig_AE}. A channel dropout vector $\boldsymbol{d}=(d_1,d_2,\ldots,d_n)$, $\  d_i \in \{0,1\},$ is a binary vector of length $n$ whose zero entries (represented by empty circles in the dropout block in Fig.~\ref{Fig_AE}) designate noise layer neurons on which the dropout is applied, i.e., whose output values are set to zero \cite{Hinton_2014}. To address the erasure channel models with multiple states ($L > 1$), we define a sequence of dropout vectors $\boldsymbol{d}_{\ell}, \ell \in \{1,2,\ldots,L\}$, where  dropout vector $\boldsymbol{d}_{\ell}$ corresponds to the $\ell$-th class. The number of zeros and ones and their positions or statistical occurrence in the dropout vectors $\{\boldsymbol{d}_{\ell}\}$  are defined as part of the rateless AE code design process. 

 In the training process, we apply the randomized dropout strategy, where different dropout vectors are applied randomly on a batch-by-batch basis. For each training batch, we first randomly sample a dropout class $\ell \in \{1,2,\ldots,L\}$ following the dropout class probability distribution $\boldsymbol{q}$, and then apply the dropout vector $\boldsymbol{d}_{\ell}$ corresponding to the dropout class $\ell$ for all the training samples in the batch. The details of applying dropout vectors on individual training batches differ for each of the erasure channel models defined in Section~\ref{MC_SC}, as we detail next.
     
\textbf{Model 1.} The dropout vector $\boldsymbol{d}_{\ell}$ for the $\ell$-th class
is constructed such that its first $r_{\ell}$ positions (corresponding to the topmost neurons in the layer) are set to one, while the remaining ones are set to zero (recall that only the first $r_{\ell}$ symbols survive the channel unerased), as shown in Fig. \ref{Fig_AE} (dropout block a)).  Note that each dropout vector $\boldsymbol{d}_{\ell}$ is fixed in advance, i.e., it is deterministic.
     
\textbf{Model 2:} The dropout vector $\boldsymbol{d}_{\ell}$ for the $\ell$-th class (characterized by the channel erasure probability $\epsilon_{\ell}$, see Section~\ref{MC_SC})
is constructed so that each of its positions is randomly and independently set to zero with  probability $\epsilon_{\ell}$, as shown in Fig. \ref{Fig_AE} (dropout block b), empty circles represent zero neurons). Note that each dropout vector $\boldsymbol{d}_{\ell}$ is now randomized, i.e., the realization of $\boldsymbol{d}_{\ell}$ is random and in general different across the training batches.

\subsection{Connections to Rateless and Rate-Compatible Codes}

The proposed AE codes provide a graceful degradation of the error probability as  additional codeword symbols are received (see Sec. \ref{results}). This behavior is akin to rateless codes, justifying the nickname \emph{rateless AE codes}. However, strictly speaking, the proposed codes are not rateless in the sense that an arbitrary number of codeword symbols can be generated from the source message. In the context of latency-constrained communications, one can set the codeword length $n$ to a sufficiently large value, e.g., to the value which corresponds to the maximum allowable decoding delay. 

From the receiver perspective, rateless AE codes allow flexible selection of the codeword length. In this sense, the proposed codes are related to rate-compatible codes for incremental redundancy hybrid automatic repeat request (HARQ) \cite{Declercq_2014}. Note also that, for any selected codeword length, we are interested in decoding the complete source message. This is in contrast to optimizing the intermediate performance of rateless codes that targets recovery of a part of the message if an insufficient number of codeword symbols are received \cite{Sanghavi_2007}. It is also different from unequal error protection  codes that provide certain messages or parts of a message a higher probability of reconstruction, as we recently investigated in the context of AE-based code design \cite{Ninkovic_2021}. 

\section{Performance Evaluation of Rateless AE Codes}
\label{results}

\label{p_7_4}


The goal of this work is to devise a code design that is able to trade off decoding delay against decoding error probability. To the best of our knowledge, code design for dying channels has only been addressed  for the binary erasure channel (BEC) \cite{Gu_2017, Zeng_2012}. Thus despite their apparent importance,  explicit constructions of codes for channel models 1 and 2 above are missing. In this work, we use an AE-based approach to address this problem.

We present results for the $(n,k)=(24,12)$ code scenario, where $M=4096$ messages are transmitted over $n=24$ channel uses. We consider a cascaded AWGN and erasure channel as described in Section~ \ref{MC_SC}.

\subsection{Rateless AE Architecture and Training Procedure}
The same training process as for the conventional AE \cite{OShea_2017}, apart from the introduction of appropriate dropout strategies, is preserved for the rateless AE training. More precisely, the rateless AE is optimized by using stochastic gradient descent with the Adam optimizer \cite{adam}. The learning rate is $\alpha=0.001$, $\beta_{1}=0.9$ and $\beta_{2}=0.999$. The training is performed at $E_b/N_0=1$ dB. For both the  conventional and rateless AE design, we apply the average power constraint across the set of codewords in the codebook. This type of normalization can be accomplished by introducing a batch normalization layer in the AE architecture (Fig. \ref{Fig_AE}) \cite{Letizia_2021}. For both the conventional and rateless AE architectures, we consider  a single fully--connected hidden layer ($H=1$) with $500$ neurons, i.e.,  $\boldsymbol{h}=h_1=500$, and a batch size $500$. Both training and test data sets are created by sampling a message set $\mathcal{M}$ uniformly at random. The training and test data set consist of \num{e5} and \num{e6} messages, respectively.

\begin{figure}[t]
\centering
\begin{tikzpicture}
  	\begin{semilogyaxis}[width=1\columnwidth, height=7.5cm, 
	legend style={at={(0.29,0.56)}, anchor= north,font=\scriptsize, legend style={nodes={scale=0.7, transform shape}}},
   	legend cell align={left},
	legend columns=1,   	 
   	x tick label style={/pgf/number format/.cd,
   	set thousands separator={},fixed},
   	y tick label style={/pgf/number format/.cd,fixed, precision=2, /tikz/.cd},
   	xlabel={$E_b/N_0$ [dB]},
   	ylabel={BLER},
   	label style={font=\footnotesize},
   	grid=major,   	
   	xmin = 1, xmax = 7,
   	ymin=0.000005, ymax=1,
   	line width=0.8pt,
   	tick label style={font=\footnotesize},]
   	\addplot[blue, mark=square] 
   	table [x={x}, y={y}] {./Tikz/Scen1(24_12)/LD_SNR/ld_4_conv.txt};
   	\addlegendentry{C-AE - $r_{\ell}=15$}
   	\addplot[dashed,blue, mark=o, mark options={solid}]
   	table [x={x}, y={y}] {./Tikz/Scen1(24_12)/LD_SNR/ra_ld4_08_005.txt};
   	\addlegendentry{R-AE $[0.8, 0.1, 0.05, 0.05]$ - $r_{\ell}=15$}
   	\addplot[dash dot,blue, mark=triangle, mark options={solid}] 
   	table [x={x}, y={y}] {./Tikz/Scen1(24_12)/LD_SNR/ra_ld4_025.txt};
   	\addlegendentry{R-AE $[0.25, 0.25, 0.25, 0.25]$ - $r_{\ell}=15$}
   	\addplot[dotted,blue, mark=star, mark options={solid}]
   	table [x={x}, y={y}] {./Tikz/Scen1(24_12)/LD_SNR/ra_ld4_005_08.txt};
   \addlegendentry{R-AE $[0.05, 0.05, 0.1, 0.8]$ - $r_{\ell}=15$}
   	
   	\addplot[red, mark=square] 
   	table [x={x}, y={y}] {./Tikz/Scen1(24_12)/LD_SNR/ld_6_conv.txt};
   	\addlegendentry{C-AE - $r_{\ell}=21$}
   	\addplot[dashed,red, mark=o, mark options={solid}]
   	table [x={x}, y={y}] {./Tikz/Scen1(24_12)/LD_SNR/ra_ld6_08_005.txt};
   	\addlegendentry{R-AE $[0.8, 0.1, 0.05, 0.05]$ - $r_{\ell}=21$}
   	\addplot[dash dot,red, mark=triangle, mark options={solid}] 
   	table [x={x}, y={y}] {./Tikz/Scen1(24_12)/LD_SNR/ra_ld6_025.txt};
   	\addlegendentry{R-AE $[0.25, 0.25, 0.25, 0.25]$ - $r_{\ell}=21$}
   	\addplot[dotted,red, mark=star, mark options={solid}] 
   	table [x={x}, y={y}] {./Tikz/Scen1(24_12)/LD_SNR/ra_ld6_005_08.txt};
   \addlegendentry{R-AE $[0.05, 0.05, 0.1, 0.8]$ - $r_{\ell}=21$}
   	
   	   	\addplot[green, mark=square] 
   	table [x={x}, y={y}] {./Tikz/Scen1(24_12)/LD_SNR/ld_7_conv.txt};
   	\addlegendentry{C-AE - $r_{\ell}=24$}
   	\addplot[dashed,green, mark=o, mark options={solid}]
   	table [x={x}, y={y}] {./Tikz/Scen1(24_12)/LD_SNR/ra_ld7_08_005.txt};
   	\addlegendentry{R-AE $[0.8, 0.1, 0.05, 0.05]$ - $r_{\ell}=24$}
   	\addplot[dash dot,green, mark=triangle, mark options={solid}] 
   	table [x={x}, y={y}] {./Tikz/Scen1(24_12)/LD_SNR/ra_ld7_025.txt};
   	\addlegendentry{R-AE $[0.25, 0.25, 0.25, 0.25]$ - $r_{\ell}=24$}
   	\addplot[dotted,green, mark=star, mark options={solid}] 
   	table [x={x}, y={y}] {./Tikz/Scen1(24_12)/LD_SNR/ra_ld7_005_08.txt};
   	\addlegendentry{R-AE $[0.05, 0.05, 0.1, 0.8]$ - $r_{\ell}=24$}

 	\end{semilogyaxis}
	\end{tikzpicture}
	\vspace*{-7mm}
\caption{Rateless AE (R-AE) versus Conventional AE (C-AE) decomposed BLER performances for different erasure channel state distributions $\boldsymbol{p}$ (Model 1, $(n,k)=(24,12)$).}
\label{Fig_3R}
\end{figure}
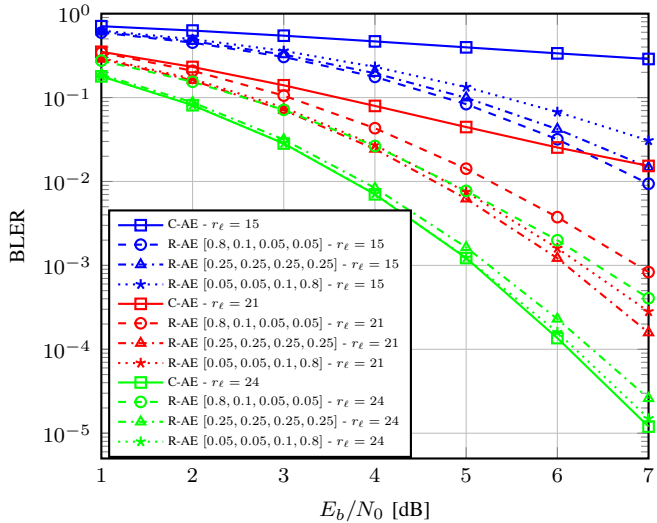

\subsection{Model 1 -- \textit{Channel with Tail Erasures}} 
We consider the tail erasure model with $L=4$ states. In each state, a receiver is able to receive the first $\boldsymbol{r}=\{15, 18, 21, 24\}$ consecutive channel symbols, respectively. The erasure channel state distribution is $\boldsymbol{p}=\{p_1,p_2,p_3,p_4\}$ (whose numerical values are specified later). Assuming that $\boldsymbol{p}$ and $\boldsymbol{r}$ are known at the transmitter, the channel dropout distribution $\boldsymbol{q}$ and vectors $\boldsymbol{d}_{\ell}$ are constructed to match the tail erasure channel parameters $\boldsymbol{p}$ and $\boldsymbol{r}$. In other words, we set $\boldsymbol{q}=\boldsymbol{p}$ and, using $\boldsymbol{r}$, we define how many neurons starting from the topmost will survive the dropout for each dropout vector class $\boldsymbol{d}_{\ell}$. Using the above parameters, we compared the rateless AE design to the conventional AE design.
 
In Fig.~\ref{Fig_3R}, we present block error-rate (BLER) results of rateless AE codes as a function of  $E_b/N_0$ for erasure channel state distributions $\boldsymbol{p}^{(1)}=\{0.8,0.1,0.05,0.05\}$, $\boldsymbol{p}^{(2)}=\{0.25,0.25,0.25,0.25\}$ and $\boldsymbol{p}^{(3)}=\{0.05,0.05,0.1,0.8\}$. For example, for the channel with state distribution $\boldsymbol{p}^{(1)}$, the probability that first $r_1=15$ symbols are received is $p_1=0.8$, the first $r_2=18$ symbols are received is $p_2=0.1$, and so on. The BLER results are presented as follows: for a different number of received symbols $r_1=15$, $r_3=21$ and $r_4=24$ (note that we skip $r_2=18$ for the sake of figure clarity), separate BLER curves are presented for each erasure channel state distribution $\boldsymbol{p}^{(1)}$, $\boldsymbol{p}^{(2)}$ and $\boldsymbol{p}^{(3)}$. We compare the BLER results with the conventional AE codes under the same settings. One can note clear performance loss of conventional AE codes as long as they do not receive all $n$ channel output symbols, unlike rateless AE codes whose BLER values degrade more gracefully. From Fig. \ref{Fig_3R}, influence of different dropout class distributions $\boldsymbol{q}$ (matched to different channel erasure state distribution $\boldsymbol{p}$) can be clearly observed. For example, if $\boldsymbol{q}=\boldsymbol{p}^{(1)}$ is applied during the training (which favors dropout vector $\boldsymbol{d}_1$ with the first $r_1=15$ ones), such a code will naturally demonstrate the best BLER performance after receiving the first 15 symbols. In contrast, if trained for the probability distribution $\boldsymbol{q}=\boldsymbol{p}^{(3)}$ that favors reception of all $n$ symbols, the rateless AE performance after $r_{\ell}=n=24$ symbols is comparable to the conventional AE BLER trained for reception of the complete codeword, while still providing significant BLER improvement for lower $r_{\ell}$ values.

\begin{figure}[t]
\centering
	\begin{tikzpicture}
  	\begin{semilogyaxis}[width=1\columnwidth, height=7.5cm, 
	legend style={at={(0.17,0.38)}, anchor= north,font=\scriptsize, legend style={nodes={scale=0.9, transform shape}}},
   	legend cell align={left},
	legend columns=1,   	 
   	x tick label style={/pgf/number format/.cd,
   	set thousands separator={},fixed},
   	y tick label style={/pgf/number format/.cd,fixed, precision=2, /tikz/.cd},
   	xlabel={$r$},
   	ylabel={BLER},
   	label style={font=\footnotesize},
   	grid=major,   	
   	xmin = 4, xmax = 13,
   	ymin=0.00001, ymax=1,
   	line width=0.8pt,
   	xtick={4, 5, 6, 7, 8, 9, 10, 11, 12, 13},
   	xticklabels={$15$, $16$, $17$, $18$, $19$, $20$, $21$, $22$, $23$, $24$},   	
   	tick label style={font=\footnotesize},]
   	\addplot[blue, mark=square] 
   	table [x={x}, y={y}] {./Tikz/Scen1(24_12)/LD_BER/c3db.txt};
   	\addlegendentry{C-AE [3 dB]}
   	\addplot[dashed,blue, mark=square, mark options={solid}]
   	table [x={x}, y={y}] {./Tikz/Scen1(24_12)/LD_BER/ra3db.txt};
   	\addlegendentry{R-AE [3 dB]}
   	
   	\addplot[red, mark=o] 
   	table [x={x}, y={y}] {./Tikz/Scen1(24_12)/LD_BER/c5db.txt};
   	\addlegendentry{C-AE [5 dB]}
   	\addplot[dashed,red, mark=o, mark options={solid}]
   	table [x={x}, y={y}] {./Tikz/Scen1(24_12)/LD_BER/ra5db.txt};
   	\addlegendentry{R-AE [5 dB]}
   	
   	   	\addplot[green, mark=triangle] 
   	table [x={x}, y={y}] {./Tikz/Scen1(24_12)/LD_BER/c7db.txt};
   	\addlegendentry{C-AE [7 dB]}
   	\addplot[dashed,green, mark=triangle, mark options={solid}]
   	table [x={x}, y={y}] {./Tikz/Scen1(24_12)/LD_BER/ra7db.txt};
   	\addlegendentry{R-AE [7 dB]}

 	\end{semilogyaxis}
	\end{tikzpicture}
	\vspace*{-6mm}
\caption{R-AE versus C-AE BLER performances as a function of the number of received symbols (Model 1, $(n,k)=(24,12)$).}
\label{Fig_4R}
\end{figure}
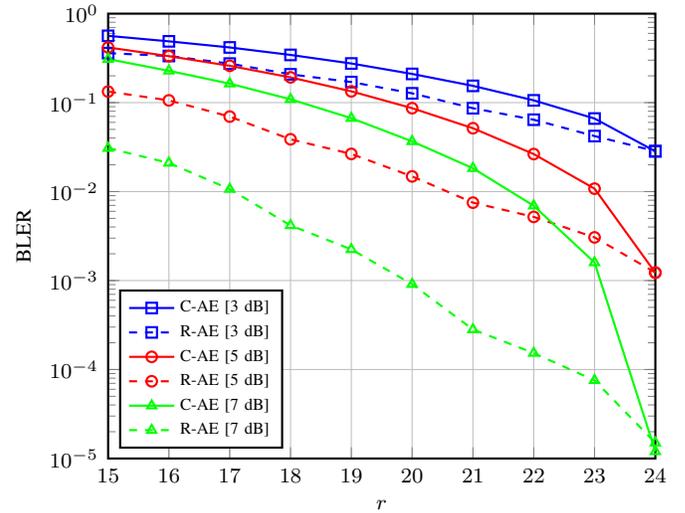

Fig. \ref{Fig_4R} illustrates the BLER performance for the rateless AE design matched to the channel state distribution $\boldsymbol{p}^{(3)}$ (i.e., $\boldsymbol{q}=\boldsymbol{p}^{(3)}$ and $\boldsymbol{d}_{\ell} = \boldsymbol{r}$) as a function of the number of received symbols $r$ for three different $E_b/N_0$ values ($3$dB, $5$dB and $7$dB). Compared to the BLER curves of the conventional AE codes, graceful degradation of BLER curves of the rateless AE codes with the increase of the number of received symbols is clearly observed. Note also that the  ability to generalize of the proposed approach is satisfactory, as from Fig. \ref{Fig_4R}, only a slight performance degradation is observed for  rates that are not seen during the training process (when $r$ is different than 15, 18, 21 and 24).

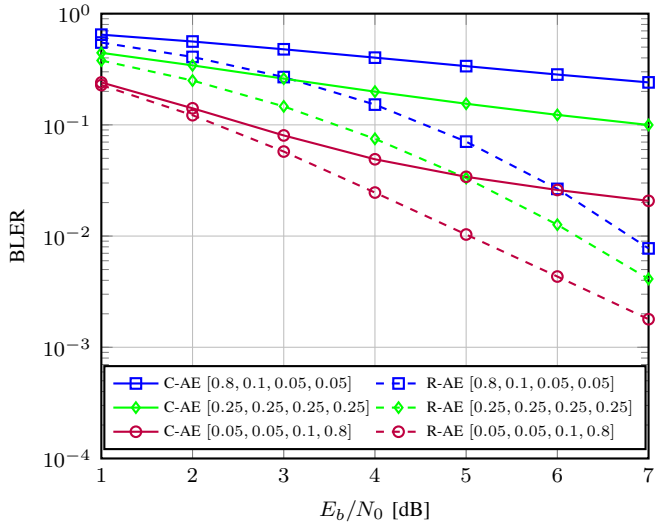
\begin{figure}[t]
\centering
	\begin{tikzpicture}
  	\begin{semilogyaxis}[width=1\columnwidth, height=7.5cm, 
	legend style={at={(0.5,0.21)}, anchor= north,font=\scriptsize, legend style={nodes={scale=0.85, transform shape}}},
   	legend cell align={left},
	legend columns=2,   	 
   	x tick label style={/pgf/number format/.cd,
   	set thousands separator={},fixed},
   	y tick label style={/pgf/number format/.cd,fixed, precision=2, /tikz/.cd},
   	xlabel={$E_b/N_0$ [dB]},
   	ylabel={BLER},
   	label style={font=\footnotesize},
   	grid=major,   	
   	xmin = 1, xmax = 7,
   	ymin=0.0001, ymax=1,
   	line width=0.8pt,
   	tick label style={font=\footnotesize},]
   	\addplot[blue, mark=square] 
   	table [x={x}, y={y}] {./Tikz/Scen1(24_12)/Avg/avg_conv_08_005.txt};
   	\addlegendentry{C-AE $[0.8, 0.1, 0.05, 0.05]$}
   	\addplot[dashed,blue, mark=square, mark options={solid}] 
   	table [x={x}, y={y}] {./Tikz/Scen1(24_12)/Avg/avg_ra_08_005.txt};
   	\addlegendentry{R-AE $[0.8, 0.1, 0.05, 0.05]$}
   	
   	
   	\addplot[green, mark=diamond] 
   	table [x={x}, y={y}] {./Tikz/Scen1(24_12)/Avg/avg_conv_025.txt};
   	\addlegendentry{C-AE $[0.25, 0.25, 0.25, 0.25]$}
   	\addplot[dashed, green, mark=diamond, mark options={solid}] 
   	table [x={x}, y={y}] {./Tikz/Scen1(24_12)/Avg/avg_ra_025.txt};
   	\addlegendentry{R-AE $[0.25, 0.25, 0.25, 0.25]$}
   	
   	
   	\addplot[purple, mark=o] 
   	table [x={x}, y={y}] {./Tikz/Scen1(24_12)/Avg/avg_conv_005_08.txt};
   	\addlegendentry{C-AE $[0.05, 0.05, 0.1, 0.8]$}
   	\addplot[dashed,purple, mark=o, mark options={solid}]
   	table [x={x}, y={y}] {./Tikz/Scen1(24_12)/Avg/avg_ra_005_08.txt};
   	\addlegendentry{R-AE $[0.05, 0.05, 0.1, 0.8]$}

 	\end{semilogyaxis}
	\end{tikzpicture}
	\vspace*{-6mm}
\caption{R-AE versus C-AE averaged BLER performances (Model 1, $(n,k)=(24,12)$).}
\label{Fig_1R}
\end{figure}

In Fig. \ref{Fig_1R}, we present the average BLER curves (averaged across the probability distribution $\boldsymbol{p}$ of the channel erasure states) versus $E_b/N_0$ (dB) for three different erasure channel state distributions $\boldsymbol{p}^{(1)}$, $\boldsymbol{p}^{(2)}$ and $\boldsymbol{p}^{(3)}$. We note that, by applying an appropriate (i.e., matched) dropout strategy for the rateless AE design, a significant improvement of the average BLER for a given erasure channel state distribution $\boldsymbol{p}$ can be achieved as compared to the conventional AE design. 

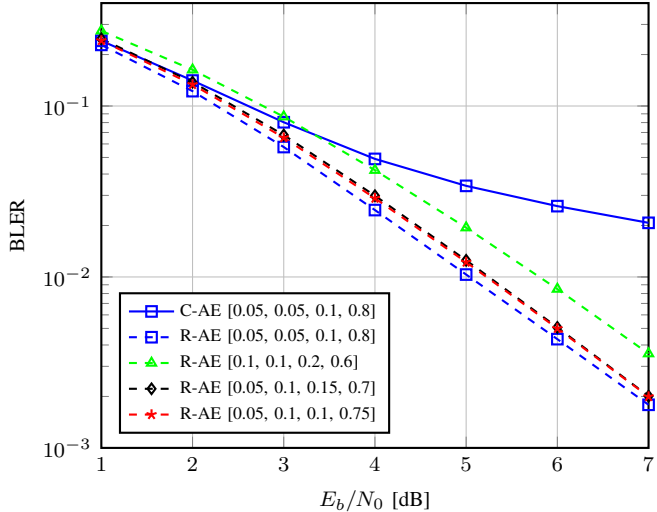
\begin{figure}[t]
\centering
	\begin{tikzpicture}
  	\begin{semilogyaxis}[width=1\columnwidth, height=7.5cm, 
	legend style={at={(0.28,0.35)}, anchor= north,font=\scriptsize, legend style={nodes={scale=0.95, transform shape}}},
   	legend cell align={left},
	legend columns=1,   	 
   	x tick label style={/pgf/number format/.cd,
   	set thousands separator={},fixed},
   	y tick label style={/pgf/number format/.cd,fixed, precision=2, /tikz/.cd},
   	xlabel={$E_b/N_0$ [dB]},
   	ylabel={BLER},
   	label style={font=\footnotesize},
   	grid=major,   	
   	xmin = 1, xmax = 7,
   	ymin=0.001, ymax=0.4,
   	line width=0.8pt,
   	tick label style={font=\footnotesize},]
   	\addplot[blue, mark=square] 
   	table [x={x}, y={y}] {./Tikz/Scen1(24_12)/Missmatch/c_005_08};
   	\addlegendentry{C-AE [0.05, 0.05, 0.1, 0.8]}
   	\addplot[dashed,blue, mark=square, mark options={solid}]
   	table [x={x}, y={y}] {./Tikz/Scen1(24_12)/Missmatch/org_005_08};
   	\addlegendentry{R-AE [0.05, 0.05, 0.1, 0.8]}
   	
   	\addplot[dashed,green, mark=triangle, mark options={solid}]
   	table [x={x}, y={y}] {./Tikz/Scen1(24_12)/Missmatch/ra_01_06};
   	\addlegendentry{R-AE [0.1, 0.1, 0.2, 0.6]}
   	
   	\addplot[dashed,black, mark=diamond, mark options={solid}]
   	table [x={x}, y={y}] {./Tikz/Scen1(24_12)/Missmatch/ra_005_07};
   	\addlegendentry{R-AE [0.05, 0.1, 0.15, 0.7]}
   	
   	\addplot[dashed,red, mark=star, mark options={solid}]
   	table [x={x}, y={y}] {./Tikz/Scen1(24_12)/Missmatch/ra_01_04};
   	\addlegendentry{R-AE [0.05, 0.1, 0.1, 0.75]}

 	\end{semilogyaxis}
	\end{tikzpicture}
	\vspace*{-6mm}
\caption{R-AE (trained for $\boldsymbol{q}=\boldsymbol{p}^{(3)}$) versus mismatched R-AE codes (Model 1, $(n,k)=(24,12)$).}
\label{Fig_2R}
\end{figure}

The previous examples assume that the transmitter knows the erasure channel state distribution $\boldsymbol{p}$, so that it can apply the matched dropout class distribution $\boldsymbol{q}=\boldsymbol{p}$ for the training process. In order to test the robustness of rateless AE codes on the mismatch of the channel state and the dropout class distributions, in Fig. \ref{Fig_2R} we examine the average performance of the rateless AE design trained for a particular dropout class distribution $\boldsymbol{q}=\boldsymbol{p}^{(3)}$ against the same code tested over three different mismatched erasure channel state distributions (see the figure legend for details). Clearly, rateless AE design whose dropout class distribution $\boldsymbol{q}$ is matched to the channel state distribution $\boldsymbol{p}$ outperforms the case where the same code is applied over the (slightly) mismatched erasure channel state distribution $\boldsymbol{p}$. On the other hand, even with mismatch, the proposed rateless AE design significantly outperforms the conventional AE design, demonstrating the inherent robustness of the rateless AE design.

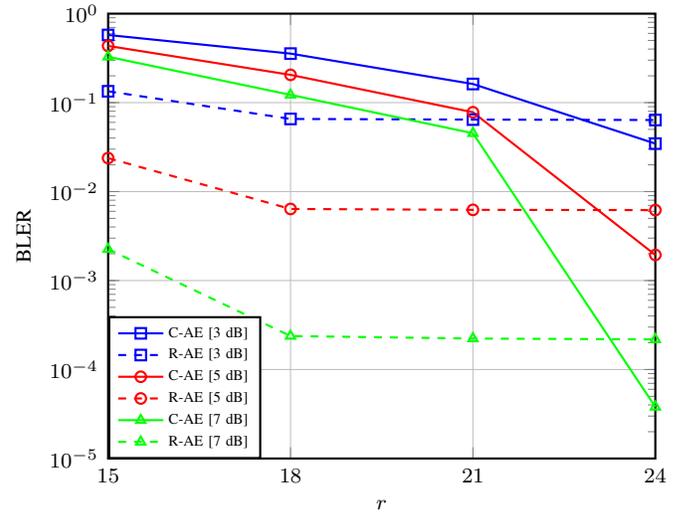
\begin{figure}[t]
\centering
	\begin{tikzpicture}
  	\begin{semilogyaxis}[width=1\columnwidth, height=7.5cm, 
	legend style={at={(0.14,0.32)}, anchor= north,font=\scriptsize, legend style={nodes={scale=0.8, transform shape}}},
   	legend cell align={left},
	legend columns=1,   	 
   	x tick label style={/pgf/number format/.cd,
   	set thousands separator={},fixed},
   	y tick label style={/pgf/number format/.cd,fixed, precision=2, /tikz/.cd},
   	xlabel={$r$},
   	ylabel={BLER},
   	label style={font=\footnotesize},
   	grid=major,   	
   	xmin = 4, xmax = 7,
   	ymin=0.00001, ymax=1,
   	line width=0.8pt,
   	xtick={4, 5, 6, 7},
   	xticklabels={$15$, $18$, $21$, $24$},   	
   	tick label style={font=\footnotesize},]
   	\addplot[blue, mark=square] 
   	table [x={x}, y={y}] {./Tikz/Scen1/Norm_ld_ber/c3db.txt};
   	\addlegendentry{C-AE [3 dB]}
   	\addplot[dashed,blue, mark=square, mark options={solid}]
   	table [x={x}, y={y}] {./Tikz/Scen1/Norm_ld_ber/ra3db.txt};
   	\addlegendentry{R-AE [3 dB]}
   	
   	\addplot[red, mark=o] 
   	table [x={x}, y={y}] {./Tikz/Scen1/Norm_ld_ber/c5db.txt};
   	\addlegendentry{C-AE [5 dB]}
   	\addplot[dashed,red, mark=o, mark options={solid}]
   	table [x={x}, y={y}] {./Tikz/Scen1/Norm_ld_ber/ra5db.txt};
   	\addlegendentry{R-AE [5 dB]}
   	
   	   	\addplot[green, mark=triangle] 
   	table [x={x}, y={y}] {./Tikz/Scen1/Norm_ld_ber/c7db.txt};
   	\addlegendentry{C-AE [7 dB]}
   	\addplot[dashed,green, mark=triangle, mark options={solid}]
   	table [x={x}, y={y}] {./Tikz/Scen1/Norm_ld_ber/ra7db.txt};
   	\addlegendentry{R-AE [7 dB]}

 	\end{semilogyaxis}
	\end{tikzpicture}
	\vspace*{-6mm}
\caption{R-AE versus C-AE BLER performances as a function of the number of received symbols - Fixed power constraint (Model 1, $(n,k)=(24,12)$).}
\label{Fig_4R_norm}
\end{figure}  
    
Finally, we emphasize the importance of the power constraint selection. Recall that, herein, we apply the average power constraint, in contrast to the fixed power constraint applied in \cite{OShea_2017}. In Fig. \ref{Fig_4R_norm}, we present the rateless AE vs the conventional AE BLER performance under the fixed power constraint for the dropout class distribution $\boldsymbol{q}=\boldsymbol{p}^{(2)}$ as a function of the number of received symbols. Besides significant degradation of the conventional AE performance unless all $n=24$ symbols are received, we note a different behavior of the rateless AE codes that essentially lose their rateless property. In other words, the rateless AE design is now optimized for a specific number of received symbols $r$ which is smaller than the codeword length $n$ (in our example, $r_2=18$) and reception of additional symbols does not improve the BLER performance.

\subsection{Model 2 -- Channel with Random Erasures}
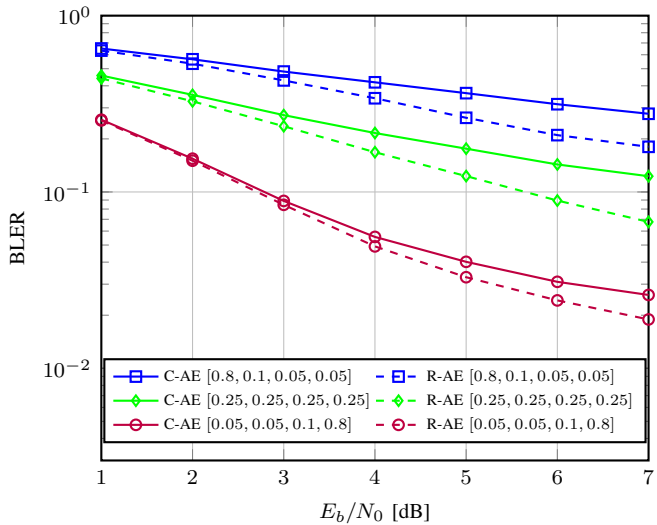
\begin{figure}[t]
\centering
	\begin{tikzpicture}
  	\begin{semilogyaxis}[width=1\columnwidth, height=7.5cm, 
	legend style={at={(0.5,0.23)}, anchor= north,font=\scriptsize, legend style={nodes={scale=0.85, transform shape}}},
   	legend cell align={left},
	legend columns=2,   	 
   	x tick label style={/pgf/number format/.cd,
   	set thousands separator={},fixed},
   	y tick label style={/pgf/number format/.cd,fixed, precision=2, /tikz/.cd},
   	xlabel={$E_b/N_0$ [dB]},
   	ylabel={BLER},
   	label style={font=\footnotesize},
   	grid=major,   	
   	xmin = 1, xmax = 7,
   	ymin=0.003, ymax=1,
   	line width=0.8pt,
   	tick label style={font=\footnotesize},]
   	\addplot[blue, mark=square] 
   	table [x={x}, y={y}] {./Tikz/Scen3/c_08_005.txt};
   	\addlegendentry{C-AE $[0.8, 0.1, 0.05, 0.05]$}
   	\addplot[dashed,blue, mark=square, mark options={solid}] 
   	table [x={x}, y={y}] {./Tikz/Scen3/ra_08_005.txt};
   	\addlegendentry{R-AE $[0.8, 0.1, 0.05, 0.05]$}
   	
   	
   	\addplot[green, mark=diamond] 
   	table [x={x}, y={y}] {./Tikz/Scen3/c_025.txt};
   	\addlegendentry{C-AE $[0.25, 0.25, 0.25, 0.25]$}
   	\addplot[dashed, green, mark=diamond, mark options={solid}] 
   	table [x={x}, y={y}] {./Tikz/Scen3/ra_025.txt};
   	\addlegendentry{R-AE $[0.25, 0.25, 0.25, 0.25]$}
   	
   	
   	\addplot[purple, mark=o] 
   	table [x={x}, y={y}] {./Tikz/Scen3/c_005_08.txt};
   	\addlegendentry{C-AE $[0.05, 0.05, 0.1, 0.8]$}
   	\addplot[dashed,purple, mark=o, mark options={solid}]
   	table [x={x}, y={y}] {./Tikz/Scen3/ra_005_08.txt};
   	\addlegendentry{R-AE $[0.05, 0.05, 0.1, 0.8]$}

 	\end{semilogyaxis}
	\end{tikzpicture}
	\vspace*{-6mm}
	\caption{R-AE versus C-AE averaged BLER performances (Model 2, $(n,k)=(24,12)$).}
\label{Fig_6R}
\end{figure}

In Fig. \ref{Fig_6R}, we present the average BLER vs $E_b/N_0$ (dB) results (averaged across different channel erasure states) for $L=4$ channel erasure states defined by the erasure probabilities $\boldsymbol{\epsilon}=\{3/8, 2/8, 1/8, 0\}$ for three different channel erasure state distributions $\boldsymbol{p}^{(1)}=\{0.8,0.1,0.05,0.05\}$ (i.e., during the training process, channel erasure probability $\epsilon_{1}=3/8$ will be applied with probability $p_1=0.8$, $\epsilon_{2}=2/8$ with probability $p_2=0.1$, and so on), $\boldsymbol{p}^{(2)}=\{0.25,0.25,0.25,0.25\}$ and $\boldsymbol{p}^{(3)}=\{0.05,0.05,0.1,0.8\}$. As expected, learning the rateless AE code for random erasures is more challenging than for the case of tail erasures, however, the proposed rateless AE design still outperforms the conventional one.

\section{Conclusion}
\label{conc}

We presented rateless AE codes, a novel class of AE codes that trade off decoding delay and reliability. By integrating a randomized dropout technique into the AE-based code design, rateless AE codes provide a graceful degradation of the decoding error probability as a function of the number of received codeword symbols. Rateless AE codes can be tailored for so-called dying channels, where the receiver observes an incomplete noisy codeword interrupted at a given (random) symbol. Such dying channels are relevant in several short-packet wireless communication scenarios. Numerical results demonstrate that the proposed rateless AE codes provide high flexibility in shaping a desired delay vs reliability behavior.     
\balance

\end{document}